\documentclass[11pt]{article}  
\usepackage{graphicx}
\newcommand{\be}{\begin{equation}}
\newcommand{\ee}{\end{equation}}
\newcommand{\bea}{\begin{eqnarray}}
\newcommand{\eea}{\end{eqnarray}}

\begin{document}
\vspace{.5in} 
\begin{center} 
{\LARGE{\bf Wormhole as a Waveguide: Case of Quantum Particles with Zero Angular Momentum}} 
\end{center} 

\vspace{.3in}
\begin{center} 
{{\bf Victor Atanasov$^{(1)}$, Rossen Dandoloff$^{(1)}$ and Avadh Saxena$^{(2)}$}}\\ 
{$^{(1)}$Department of Condensed Matter Physics and Microelectronics, Faculty of Physics, 
Sofia University, 5 Blvd. J. Bourchier, 1164 Sofia, Bulgaria \\ 
$^{(2)}$Theoretical Division and Center for Nonlinear Studies, Los
Alamos National Laboratory, Los Alamos, New Mexico 87545, USA}
\end{center} 

\vspace{.9in}
{\bf {Abstract:}}  
We consider a static wormhole as a waveguide and determine the conditions for full transmission through the wormhole waveguide for a quantum particle with zero angular momentum. We find that the waveguide is transparent when the de Broglie wavelength of the quantum particle is an integer times twice the throat diameter of the wormhole. 
Such an effect may be realizable in graphene, plasmonic or optical wormholes.

\newpage 
  
\section{Introduction}
There have been recent investigations of quantum mechanics on different curved surfaces, among others on a catenoid \cite{bjorn} which represents a two-dimensional (2D) section of a wormhole \cite{thorne}. There, 
for particles with zero angular momentum and zero energy, the quantum potential created by the geometry becomes transparent. The static wormhole has also been studied in connection with the anticentrifugal 
potential \cite{wormhole}. The geometry of the static 3D wormhole reminds us of a waveguide and here we will explore this possibility in the spirit of a study of {\it transparent} rectangular waveguides 
\cite{transparent}.  In a recent paper we have shown that a wormhole with a certain profile of the radius can exhibit anticentrifugal potential which was previously known to appear only in 2D \cite{wormhole}.  
We have also considered a generalized anticentrifugal potential \cite{generalized}. Recently a rectangular toroidal dielectric waveguide has been proposed as an experimental platform to observe quantum anticentrifugal force \cite{dielectric}. 

Realization of a wormhole in condensed matter as well as optics is a worthwhile pursuit. The behavior of massless fermions in a graphene wormhole in an external magnetic field has been studied \cite{graphene1} including its electronic properties \cite{graphene2, graphene3}. Considering wormholes as a topological feature of space, electromagnetic wormholes were proposed \cite{EM}. Following this proposal plasmonic analogs of wormholes have been studied using toroidal metamaterials \cite{plasmonic}.  In a similar vein, using magnetic metamaterials and metasurfaces magnetostatic wormholes have been experimentally demonstrated \cite{magnetic}. Traversable wormholes in condensed matter have been proposed by considering two entangled superconducting qubits coupled to a dc-SQUID array embedded in a microwave transmission line \cite{qubit}. Similarly, optical wormholes comprising liquid crystal films on a catenoid have been theoretically studied \cite{optical} and proposed using nanophotonics \cite{photonic}.  By invoking transformation optics and considering flexural elastic waves on a curved plate, even a 2D wormhole can be mimicked \cite{flexural}. Even the esoteric question of whether quantum field theory can enforce an averaged version of the `weak energy condition' has been pondered in the wormhole context \cite{morris}. 

On the other hand, we have studied quantum waveguides and transformed the problem into a particle passing (or tunneling) through a potential barrier \cite{transparent}.  Because of its 3D geometry a static wormhole can be viewed as a waveguide and here we will investigate under what conditions such a waveguide would be transparent to a quantum particle.  This would add yet another peculiar property of particles 
with zero angular momentum. We note here that it is known that a particle passing through a waveguide acquires a geometric quantum phase \cite{leblond}. Next, we consider two 
situations depending on whether the de Broglie wavelength of the particle is larger or smaller compared to the wormhole size. 


\begin{figure}[t]
\begin{center}
\includegraphics[scale=0.5]{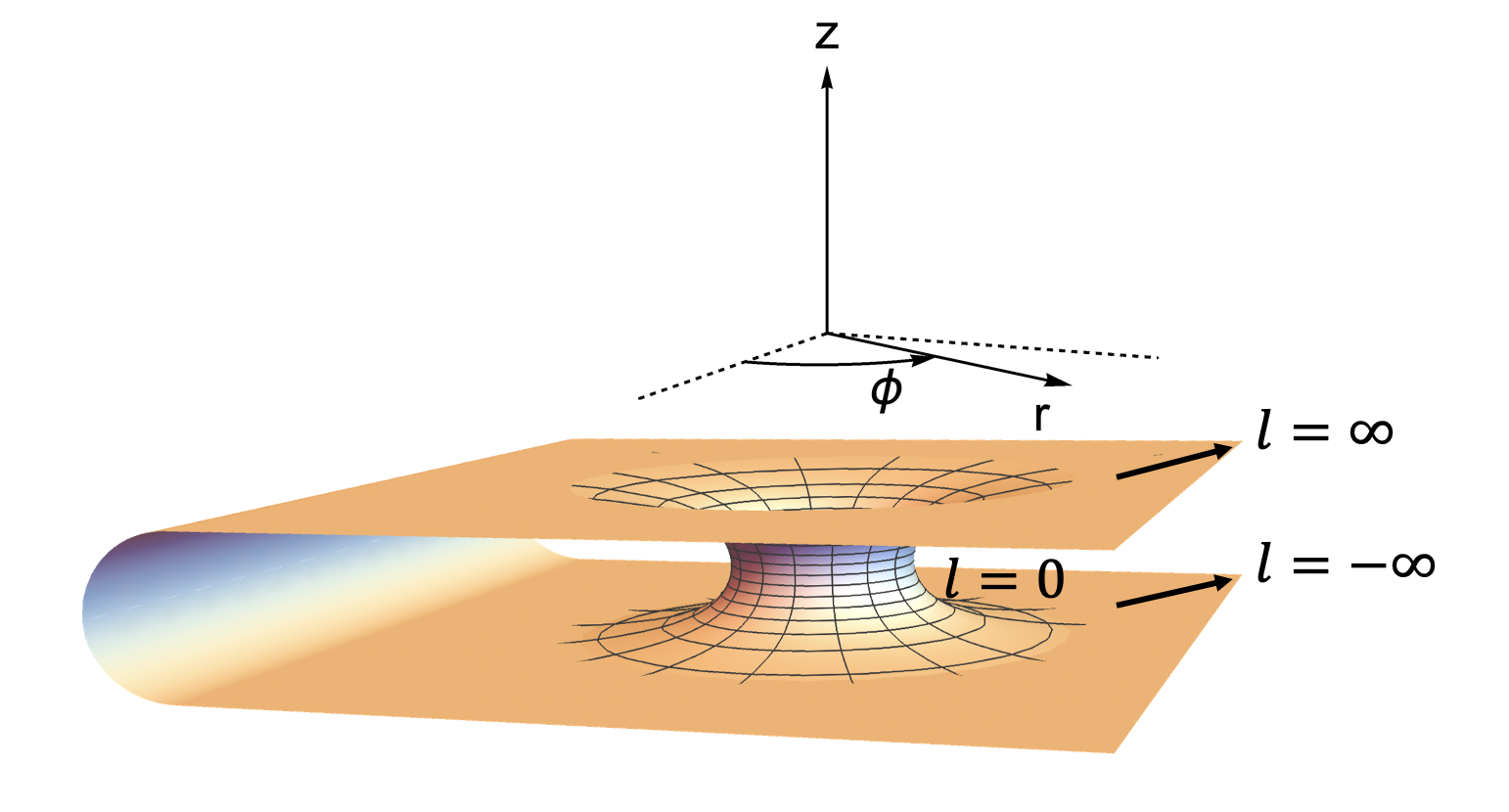}
\caption{\label{fig} Schematic diagram of a wormhole and the associated cylindrical coordinate system at a particular time with the external universe embedded in higher-dimensional space.  The throat radius is $b_0$. }
\end{center}
\end{figure} 

\section{Resonances: de Broglie wavelengths larger than the wormhole} 
We start by considering the line element for a {\it static} wormhole which is given by \cite{thorne, wormhole}  
\be 
ds^2 = dr^2 + (b_0^2 + r^2) (d\theta^2 + \sin^2\theta d\phi^2) \,, 
\ee 
where $b_0$ denotes the radius of the throat of the wormhole and $r$ is the radial distance from 
the throat (see Fig. 1). Note that at $r=\infty$ the line element is simply the line element for the normal flat $R^3$. From the line element we can read off the Lam\'e coefficients which will allow us to write the effective Schr\"odinger equation (see Eq. (8) in \cite{wormhole}):
\be
-\frac{\hbar^2}{2m_0} \frac{\partial^2\Psi_L}{\partial r^2}  + V_{eff}\Psi_L = E\Psi_L \,, 
\ee
where $L$ is the angular momentum quantum number and the effective potential is given by 
\be\label{effSchr} 
V_{eff} = \frac{\hbar^2}{2m_0} 
\left[ {\frac {L \left( L+1 \right) }{{r}^{2}+{b_{{0}}}^{2}}}+{
\frac {{b_{{0}}}^{2}}{ \left( {r}^{2}+{b_{{0}}}^{2} \right) ^{2}}}
  \right] \,. 
\ee 
Note that at $r=\pm\infty$ the Schr\"odinger equation is the usual one for the flat $R^3$. Exact solution for the wave function $\Psi_L$ in Eq.~(2) with the potential in Eq.~(3) is given in the Appendix. For particles with zero angular momentum ($L=0)$ the conservation of energy when they go through 
the wormhole reads \cite{leblond} 
\be 
\frac{\hbar^2 p_0^2}{2m_0} = \frac{\hbar^2 p_1^2}{2m_0} + \frac{\hbar^2}{2m_0} \frac{b_0^2}{(b_0^2+r^2)^2} \,, 
\ee
where the l.h.s. refers to the particle energy before entering the wormhole  waveguide (the initial energy at $r=-\infty$) and the first term on 
the r.h.s. is the energy related to the new longitudinal linear momentum after entering the waveguide.  This implies that 
\be 
\frac{\hbar^2 p_0^2}{2m_0} = \frac{\hbar^2}{2m_0}(p_0 -\Delta p)^2 + \frac{\hbar^2}{2m_0} \frac{b_0^2}{(b_0^2+r^2)^2} \,, 
\ee 
where 
\be 
\Delta p = \frac{1}{p_0} \frac{b_0^2}{(b_0^2+r^2)^2} \,. 
\ee 
Here we have supposed, as in the case of simple rectangular waveguides \cite{transparent, leblond}, that $\Delta p \ll p_0$. This means that
the energy of the incoming particle has to be much bigger than the maximum of the potential at $r=0$, namely:
\be 
\frac{1}{p_0} \frac{b_0^2}{(b_0^2+r^2)^2} \ll p_0 \,, 
\ee
and for $r=0$  we have
\be
\frac{\hbar^2p_0^2}{2m_0} \gg \frac{\hbar^2}{2m_0b_0^2} \,. 
\ee

The total phase accumulated through the passage \cite{transparent} is thus given by 
\be 
\Delta\Phi = h \int_{-\infty}^{+\infty} \Delta p ~ dl = 2h\int_0^\infty \Delta p~dl = h\frac{b_0^2}{p_0} 
\int_0^\infty \frac{dl}{(b_0^2+r^2)^2} = \frac{\pi}{4} \frac{h}{b_0 p_0} \,. 
\ee 
If we want the accumulated phase to be $\Delta\Phi = n\pi$, then we have:
\be
\frac{\pi}{4} \frac{h}{b_0 p_0} = n\pi \,. 
\ee

Now, since $p_0=h/\lambda$, where $\lambda$ is the de Broglie wavelength of the quantum particle, we find that for $\lambda=4nb_0 = 2n(2b_0) = 2nd$ 
(where $d=2b_0$ is the diameter of the throat of the wormhole) the wormhole becomes a {\it transparent waveguide}. 
Let us note that as a difference, compared to usual waveguides, here there are no diffraction effects when a particle enters or exits the waveguide.

\section{de Broglie wavelengths smaller than the wormhole}

The potential (\ref{effSchr}) in the effective Schr\"odinger equation (2) has a Fourier transform $V(q)$
\begin{eqnarray}\label{}
V(q)=\left(\frac{\hbar^2}{2m_0}\right) {\frac { \pi \, {{\rm e}^{-b_{{0}} q}}\, \left( 2\,{L}  (L+1)+b_{{0}}q+1 \right) }{2 b_{{0}}}} \, ,
\end{eqnarray}
therefore we can evaluate the scattering amplitude $A = - m_0 /(2 \pi \hbar^2) \, V(q)$ in the first Born approximation (here $q^2=4 k^2 \sin^2 (\theta/2)$), namely
\begin{eqnarray}\label{}
A=-\,{\frac {  2\,{L} (L+1) +2\,b_{{0}}\sqrt {{
k}^{2} \left( \sin \left( \theta/2 \right)  \right) ^{2}}+1
  }{8 b_{{0}}}} {{\rm e}^{-2\,b_{{0}}\sqrt {{k}^{2} \left( \sin \left( 
\theta/2 \right)  \right) ^{2}}}} \,. 
\end{eqnarray}
The differential scattering cross-section $d \sigma /d \Omega = |A|^2$ exhibits a constant behavior at zero energy $k=0,$ i.e. in the long wavelength limit there is no angular dependence
\begin{eqnarray}\label{}
A^2_{k=0}={\frac { \left( 2\,{L}^{2}+2\,L+1 \right) ^{2}}{64\,{b_{{0}}}^{2}}} \,, 
\end{eqnarray}
which is non-vanishing at any real valued integer angular momentum quantum number $L.$

The total scattering cross-section $\sigma = \int |A|^2 d \Omega$ is given by
\begin{eqnarray}\label{}
\frac{\sigma}{2 \pi} &=& -{\frac {{{\rm e}^{-4\,b_{{0}}k}}k}{16 b_{{0}}}}+{\frac { \left( -
64\,{L}^{2}{b_{{0}}}^{2}-64\,L{b_{{0}}}^{2}-56\,{b_{{0}}}^{2} \right) 
{{\rm e}^{-4\,b_{{0}}k}}}{512\,{b_{{0}}}^{4}}}\\
\nonumber &&+{\frac { \left( -32\,{L
}^{4}b_{{0}}-64\,{L}^{3}b_{{0}}-96\,{L}^{2}b_{{0}}-64\,Lb_{{0}}-36\,b_
{{0}} \right) {{\rm e}^{-4\,b_{{0}}k}}}{512\,{b_{{0}}}^{4}k}}\\
\nonumber &&+{\frac {
 \left( 8\,{L}^{4} +16\,{L}^{3} +24\,{L}^{2}+16+9\, \right) ({{\rm e}^{4\,b_{{0}}
k}}-1) {{\rm e}^{-4\,b_{{0}}k}}}{512\,{b_{{0}}}^{4}{k}^{2}}} \,, 
\end{eqnarray}
which for $L=0$ is vanishing when
\begin{eqnarray}\label{}
{\frac { \left( -9-32\,{x}^{3}+9\,{{\rm e}^{4\,x}}-56\,{x}^{2}-36\,x
 \right) {{\rm e}^{-4\,x}}}{512\,{x}^{2}{b_{{0}}}^{2}}}=0 \,.
\end{eqnarray}
Here $x=b_0 k.$ Counterintuitively, $k=0$ is not a solution and the total scattering cross-section vanishes asymptotically as depicted in Fig.\ref{fig}.

\begin{figure}[t]
\begin{center}
\includegraphics[scale=0.5]{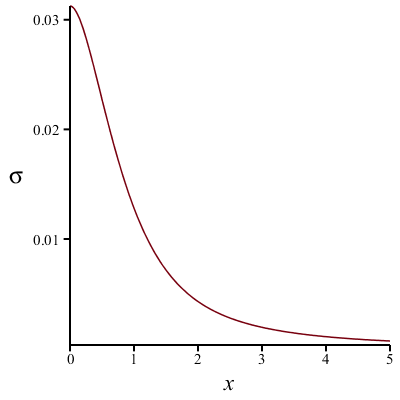}
\caption{\label{fig} The total scattering cross-section $\sigma$ for $L=0$ in the first Born approximation. Here $x=b_0 k.$}
\end{center}
\end{figure}

Note that the above result is obtained in the first Born approximation and points to a {\it reflectionless condition} for quantum particles (with vanishing angular momentum $L=0$) with de Broglie wavelengths $\lambda \ll 2 b_0 \pi$, much smaller than the diameter of the throat of the wormhole.  

The phase analysis from the previous section points to the wormhole being transparent for particles with resonance de Broglie wavelengths larger than the diameter of the throat of the wormhole, which is reminiscent of the Ramsauer-Townsend effect \cite{townsend}.  The latter relates to the scattering of low-energy electrons by atoms in noble gases.

\section{Conclusion}
Realization of the wormhole geometry  in condensed matter and optics is becoming an important pursuit. We have shown that the static wormhole geometry may serve as a quantum waveguide. 
This waveguide may turn out to be transparent for quantum particles with zero angular momentum $L=0$.  In fact, this waveguide is transparent for a whole series of de Broglie wavelengths 
that are multiples of the diameter ($2b_0$) of the throat of the wormhole. This property of the particles with $L=0$ adds to the well known peculiar anticentrifugal potential of such particles on 2D 
surfaces \cite{wormhole}. Understandably, for very small radius $b_0$ of the throat of the wormhole, the energy of the incoming particle has to be considerable. Given the recent experimental progress, a verification of our prediction 
for zero angular momentum particles passing through a graphene wormhole geometry \cite{graphene1, graphene2, graphene3}, and possibly by optical, plasmonic or other means 
\cite{EM, plasmonic, magnetic, qubit, optical, photonic, flexural}, would be highly desirable. 

\section*{Acknowledgment}  The work at Los Alamos National Laboratory was carried out under the auspices of the U.S. DOE and
NNSA under Contract No. DEAC52-06NA25396 and supported by U.S. DOE (A.S.).



\section*{Appendix: Exact solution}

The effective Schr\"odinger equation (2) with the potential (3) can be cast into the form
\begin{eqnarray}\label{heun}
{\frac {\partial^{2} \Psi_L }{{ \partial}{r}^{2}}} +{k}^{2} \Psi_L
 = \left( {\frac {L \left( L+1 \right) }{{r}^{2}+{b_{
{0}}}^{2}}}+{\frac {{b_{{0}}}^{2}}{ \left( {r}^{2}+{b_{{0}}}^{2}
 \right) ^{2}}} \right) \Psi_L \left( r \right) \,, 
\end{eqnarray}
where $k^2=2mE/\hbar^2$. Equation (\ref{heun}) has the following exact solution 
\begin{eqnarray}\label{}
\Psi_L \left( r \right) &=& {C_1}\,\sqrt {{r}^{2}+{b_{{0}}}^{2}}{
H_C} \left( 0,-\frac12,0,-\frac{{k}^{2}{b_{{0}}}^{2}}{4}, \frac{{k}^{2}{b_{{0}}}^{2}}{4} - \frac{{L}(L+1) }{4} +\frac14,-{\frac {{r}^{2}}{{b_{{0}}}^{2}}} \right) \\
\nonumber &&+
{C_2}\,r \sqrt {{r}^{2}+{b_{{0}}}^{2}} {
H_C} \left( 0,\frac12,0,-\frac{{k}^{2}{b_{{0}}}^{2}}{4}, \frac{{k}^{2}{b_{{0}}}^{2}}{4} - \frac{{L}(L+1) }{4} +\frac14,-{\frac {{r}^{2}}{{b_{{0}}}^{2}}} \right) ,
\end{eqnarray}
which holds for $r<b_0,$ i.e. in the interior of the wormhole. Here $
H_C$ is the confluent Heun function \cite{ronveaux} and $C_1$, $C_2$ are integration constants.

\end{document}